\documentstyle[12pt]{article}
\title{}
\begin{document}
\begin{center}
\begin{Large}
\baselineskip=10mm plus 1mm minus 1mm
{\bf COSMOLOGY OF VACUUM}\\
\vspace{2cm} Vladimir Burdyuzha\\
\end{Large}
\vspace{0.5cm} Astro-Space Center, Lebedev Physical Institute, Russian Academy of Sciences,\\
Profsoyuznaya 84/32, 117997 Moscow, Russian Federation\\
\vspace{1cm}
\begin{Large}
Grigoriy Vereshkov\\
\end{Large}
\vspace{0.5cm} Physical Department of Rostov State University, Stachki str.194,\\
344104 Rostov on Don, Russian Federation\\
\end{center}
\newpage
\begin{abstract}

Shortly the vacuum component of the Universe from the geometry point of view and from
the point of view of the standard model of physics of elementary particles is discussed.
Some arguments are given to the calculated value of the cosmological constant
(Zeldovich approximation). A new component of space vacuum (the gravitational
vacuum condensate) is involved the production of which has fixed time in our Universe.
Also the phenomenon of vacuum selforganization must be included in physical consideration
of the Universe evolution.
\end{abstract}
\newpage
\begin{center}
{\bf 1. WHAT IS VACUUM ?}\\
\end{center}

The refusal of the idea that vacuum is a emptiness - the conceptual statement
of modern physics.\\
{\bf In classical physics}: vacuum is the simplest system -world without particles
and this world is flat (psevdoeuclidean);\\
{\bf In quantum physics}: vacuum is a system of vacuum condensates arising in
the processes of relativistic phase transitions during early Universe evolution;\\
{\bf In geometry physics}: vacuum is a state in which geometry of space-time is
not deformed;\\
{\bf More general}: vacuum is a stable state of quantum fields without excitation
of wave modes (nonwave components are condensates).\\
Recent observations have shown that $ \Omega_{\Lambda}$ -the density of dark energy
(DE) dominates in $ \Omega_{0}$ - the total density of the Universe.

$$\Omega_{0} = \Omega_{\Lambda} + \Omega_{DM} + \Omega_{b} + \Omega_{\gamma} + \Omega_{\nu}$$

here: $\Omega_{DM}$ - the density of dark matter energy; $\Omega_{b}$ - the density of baryon component;
$\Omega_{\gamma}$ - the density of radiation and $\Omega_{\nu}$ - the density of neutrino component.

$$\Omega_{DE} \simeq 0.73 \pm 0.04; \Omega_{DM} \simeq 0.27 \pm 0.04; \Omega_{b} \simeq 0.045  \pm 0.005$$
(Starobinsky, 2004; Lahav and Liddle, 2004) \\

This DE is the reason of  accelerated expansion of the Universe after $z\sim 0.5$ (now z=0). Before the red
shift $z\sim 0.5$ our Universe have had decelerated expansion (Riess et al, 2004).
Till now we know very small about dark energy. There are some proposals as it can be.\\

\hspace{0.5cm}/ or cosmological constant ($\Lambda-term$) vacuum energy\\

{\bf Dark energy}  or quintessence (scalar field)\\

\hspace{0.5cm}$\backslash$ or phantom energy\\

\begin{center}
{\bf Difference}
\end{center}
\hspace{3cm} a) $w\equiv p/\epsilon =-1$\\
${\bf \Lambda-term}\hspace{1cm}$ b) practically constant in time \\
\hspace*{3cm} c) spatially homogeneous\\

Weinberg (1989); Turok,Hawking (1998); Burdyuzha,Vereshkov (2002); Dolgov (2004).\\
\newpage
\hspace{4cm} a)$ w\equiv p/\epsilon\neq-1 \hspace{1cm} w>-1$\\
{\bf Quintessence}\hspace{2cm} b) time evolution\\
\hspace*{4.5cm} c)spatially heterogeneous\\
\hspace*{4.5cm} d) w(z)\\

Starobinsky (1998); Armendariz-Picon et al. (2000); Hebecker, Wetterich (2000); Rubakov (2000).\\

{\bf Phantom energy} \hspace{1cm}a) $w\leq-1$\\

Vikman (2004).

\begin{center}
{\bf 2. GEOMETRIZATION OF PHYSICS}\\
\end{center}
/macroworld - Riemann geometry; Einstein GR\\
{\bf This process was initiated by M. Grossman}\\
$\backslash$ microworld - geometry of stratifications; gauge fields;\\
Kaluza, Klein, Konopleva, Popov, Hawking, Penrose\\

In Einstein theory the gravitational field is a measure of space-time curvature.
Other physical fields participate in this curvature or in the reaction on this curvature.
Electro-magnetic, weak and strong interactions are not connected with this curvature.
They are connected with a distortion of other geometry- a stratification of space-time
(3 types of interactions correspond different forms of stratifications).
The stratificated space goes from the theory of elasticity.
In this space all points are changed by new stratifications; that is points have
the interior structure! (physical model of stratificated space is a multitude of
identical but different orientated elements (Latipov et.al.2001)). A combination of
stratifications corresponding electro-magnetic interaction is able to be in a
non-deformation state. A combination of stratifications corresponding weak and
strong interactions are able to be in a deformation state (these states are
energetically profitable (confinement as the example)).
The transition from the 4-dim space to a stratificated space means that the space of
interactions may be multidimensional one. Geometry of the stratificated  space includes Riemann
geometry as a particular case because Riemann space posses properties of Minkowsky
space locally. Therefore geometry of the stratificated  space gives the possibility to consider each gauge field separately
changed the structure of a stratification in accordance with the gauge group.

{\bf Thus, the main element of our World may not be the 4-dim continuum of Einstein-Minkowsky.
It may be a 4-dim distorted and stratificated geometry. In this conception of fields vacuum is
the state in which geometry of space-time does not deform}.

Here some comparisons may be useful: for compensation of physical space curvature
Christoffel symbols are involved; for compensation of stratifications of charge spaces
gauge fields are involved. Stratificated structures of weak and strong interactions
deform spontaneously but the character of deformation is different. The structure of
stratifications of weak interaction is classically determined (zero vibrations
perturb slightly its). The structure of stratifications of strong interaction is formed
other way (Latipov et.al., 2001). These structures form the state of physical vacuum and
their name is VACUUM CONDENSATES. The properties of matter are completely defined
properties of these vacuum condensates that is all particles of matter-quarks,e,w,z obtain
their masses by means of interaction with vacuum condensates.

\begin{center}
{\bf 3. VACUUM OF THE STANDARD MODEL}\\
\end{center}

{\bf The physical vacuum is a medium having specific properties}:\\

it is a carrier of energy and it has pressure;\\
it has a interior microscopic structure;\\
it has excited states of wave and soliton types;\\
it the a medium without resistance;\\
the wave excitements are usual particles; \\
the soliton excitements are Higgs bosons. \\

{\bf Vacuum has important differences from usual medium}:\\

it is impossible to connect a definite reference system;\\
it has a specific equation of state $p=-\epsilon;$ or $p=-\epsilon(a)$;\\
it looks the same from any reference system.\\

This equation of state provides automatically constancy the density of energy
and pressure in all processes of heating and cooling. The constant density of vacuum
energy was at first appeared in Einstein equations as $\Lambda-term$.

\begin{center}
$$R_{\mu\nu} - 1/2g_{\mu\nu}R=8\pi G T_{\mu\nu} + \Lambda g_{\mu\nu}$$
\end{center}

Other words vacuum is the medium with a very complicated structure which had
changed during early Universe evolution and which can be rebuilded by change of
matter existence (the Brookhaven experiment will show this). Probably, $\Lambda-term$
may consist of some components:

\begin{center}
$$\Lambda=\Lambda_{QF} + \Lambda_{WH} + \Lambda_{G}$$
\end{center}

$\Lambda_{QF}$ is formed by zeroth vibrations of quantum fields and nonperturbative condensates;
$\Lambda_{WH}$ is formed by worm holes;
$\Lambda_{G}$ is formed by a gravitation vacuum condensate-GVC:

$$\Lambda_{G}\equiv\frac{9\pi^{2}}{2\kappa^{2}}\lambda_{n}$$

$\lambda_{n}$ defines the spectrum of GVC possible states (we shall consider this
later in details). The general for all items is they were created during the Universe evolution.
In the region of temperatures $150 MeV < T < 100 GeV$ we have:

$$\Lambda_{SM}= -\frac {m_{H}^{2} m_{W}}{2g^{2}} - \frac {1}{128\pi^{2}}(m_{H}^{4} + 3m_{z}^{4} + 6m_{W}^{4} - 12m_{t}^{4})$$

where: the first term is the density energy of a quasiclassical Higgs condensate;
the second term the polarization of vacuum of quantum fields (here boson
contributions are negative but fermion one is positive). In standard model the mutual
compensation of contributions is prohibited by the condition of stability.
Thus, the statement that vacuum energy had decreased during Universe evolution
has absolute character. The untriviality of QCD vacuum is: a medium forms nonzero values of
the gluon condensate ($F^{a}_{\mu\nu}=(600MeV)^{4}$ and the quark condensate $<qq> = -(250MeV)^{3}$.
The properties of this vacuum give the phenomenon of confinement.

\begin{center}
{\bf 4. PHENOMENON OF SELF-ORGANIZATION}\\
\end{center}

\hspace{5cm}$\Omega_{\Lambda} = 1 \Longrightarrow \Omega_{\Lambda} = 0.7$

$\Omega_{\Lambda} = (10\div20)\Omega{b}$ is the hard established fact.
\begin{center}
($\Omega_{\Lambda}\sim0.7$;\hspace{1cm}$\Omega_{m}\sim0.3$;\hspace{1cm} $\Omega_{b}\sim0.03\div0.07$)\\
\end{center}
$\rho_{\Lambda} \sim 2\cdot 10^{-8} erg/cm^{3}$ in the Universe now. How the modern density of vacuum energy
was formed in the Universe?
Vacuum in the Universe is the combination of a large number of mutual connected subystems
(a quark-gluon condensate(QGC); a Higgs condensate; a gravitational one and others).
$\rho_{QGC}\sim10^{36} erg/cm^{3}$ QGC has opposite sign to the density of vacuum energy.
$\rho_{HC}\sim 10^{55} erg/cm^{3}$ Higgs condensate has opposite sign to the density of
vacuum energy.\\
How these subsystems are coordinated? How the compensation of huge positive and negative contributions did take place?
Probably the compensation of the density of vacuum energy (if it was positive) has occured during initial
Universe evolution by negative contributions for the relativistic phase transitions.Our Universe was losing
the symmetry the next probable way:

$$P\Longrightarrow D_{4}\times[SU(5)]_{SUSY}\Longrightarrow D_{4}\times[U(1)\times SU(2)\times(SU(3)]_{SUSY}\Longrightarrow$$

$$D_{4}\times U(1)\times SU(2)\times SU(3)\Longrightarrow D_{4}\times U(1)\times SU(3)\Longrightarrow D_{4}\times U(1)$$

Of course, whole chain is our proposal but one can be sure only in two last transitions.
Any changes on one scale lead to rebuilding of vacuum condensates on other scales.
This is a quantum self-organization of vacuum. In the frame of preon scenario
(Burdyuzha et al.2004) all vacuum subsystems have a negative density of energy.
In the frame of supersymmetric scenario the possibility is to get the exact compensation
(although in the low energetic region that compensation is absent).

On this way the problem of $\Lambda-term$ is not solved and here the key moment is
a quantum theory of gravity. Gravitational vacuum (as and vacuum QCD) has a
complex quantum topological structure. If the vacuum of QCD represents the combination of
structures in stratifications of space-time then gravitational vacuum represents
the collection of topological structures in curvatures of space-time. They have the Planck size.\\
\hspace{5cm} $D=3 \Longrightarrow$ worm holes, here D is dimension;\\
\hspace{5cm} $D=2 \Longrightarrow$ micromembranes;\\
\hspace{5cm} $D=1 \Longrightarrow$ microstrings;\\
\hspace{5cm} $D=0 \Longrightarrow$ a gas of point defects (monopoles).\\
{\bf The important moment is the space filled of worm holes is a carrier of positive energy!}
That is the presence of worm holes provides the principal possibility of mutual
compensation of different contributions in total energy of vacuum. Probably it
is necessary to reject the idea of parameters fitting of vacuum subsystems in
the scale of the Universe till 120 signs at Planck energy. It is necessary to
invent a law of dynamical coordination of vacuum subsystems. This law must "govern"
vacuum (its evolution). The aim is practically zeroth $\Lambda-term$. Here we
have self-organizing and evolving vacuum.

\begin{center}
{\bf 5. COSMOLOGICAL CONSTANT AS A DYNAMICAL VARIABLE}\\
\end{center}

Of course, vacuum energy could be changed by jumps after each relativistic phase transitions
(negative contributions of condensates in vacuum energy during Universe evolution).

Other possibility is when vacuum energy could be changed by smoothly (this case has
the special name a {\bf quintessence}). In observed cosmology this assumption  has the confirmation
(bursts from far supernovae for $z>1$). Other words $\Lambda-term$ has been undergone the slow
cosmological evolution (probably $\epsilon_{vac}\sim 1/a$ might be).
The equation of state during Universe evolution may be even soften $(p = -2/3\epsilon)$.

{\bf It is a beautiful idea! But which is a mechanism of this? And why?}\\
Probably the density of vacuum energy slowly diminishing to zero is a summary result of
complex processes of vacuum subsystems self-organization on different of space time scales.
That is we can not describe geometry by the classical language and this demonstrates
very well physics of vacuum.

\begin{center}
{\bf 6. CALCULATION OF $\Lambda-TERM$}\\
\end{center}
Zel'dovich (1967) attempted to account the density energy of vacuum. He have got
a formula for this inserted in its the mass of electron or proton:

$$\Lambda = 8 \pi G^{2}m^{6}\hbar^{-4}$$

here: G is gravitational constant.

The physical meaning of this formula is gravitational forces in vacuum fluctuations
as a high order effect. The calculation has shown the agreement with observed value of
$\Lambda-term$ is not good. Kardashev (1998) proposed to modify the Zeldovich formula and
to use the mass of pions for these calculations:

$$\Lambda = 8 \pi G^{2} m_{\pi}^{6} h^{-4}$$
here: h is instead $\hbar$;

Remarkably that calculated values of $\Lambda-term$ using the Zeldovich formula gives
$\Omega_{\Lambda}\sim0.7$ if Hubble constant $H_{0}=72.5 (km/s/Mpc)$.
Here some physics takes place even (Burdyuzha,1998). For temperature of chiral
symmetry breaking ($T\sim150 MeV$) the main contribution in periodic collective
motions of a nonperturbative vacuum quark-gluon condensate carries  $\pi$ mesons  as
the lightest pseudo-Goldstone particles. Here the spectrum of excitations reflects
the properties of the ground state.
Thus, if $$ m_{\pi} = \frac{2m_{\pi^{\pm}} + m_{\pi^{0}}}{3} = 138.0387 MeV $$  then we have:\\

$H_{0}(km/s/Mpc)$60\hspace{1cm}65\hspace{1cm}70\hspace{1.3cm}75\hspace{1.5cm}80\\

$\Omega_{\Lambda}$\hspace{2.2cm}1.02\hspace{0.7cm}0.87\hspace{0.6cm}0.75\hspace{1.cm}0.65\hspace{1.2cm}0.57\\

Substituting the Planck mass into these formulae the difference in 120 orders between
the vacuum energy density and its value at the Planck time can be found. The series of
relativistic phase transitions must be continued by phase transitions in dark matter.
Of course, many questions are remained. Why do $\Lambda$ not depend on other constants? and so on....
But there are important results of these calculations of $\Lambda-term$.
In the present epoch the vacuum energy density is nonzero and positive and vacuum
is in a "excited state". Besides, we "live" in this cosmological quark-gluon vacuum.

\begin{center}
{\bf 7. GRAVITATIONAL VACUUM STRUCTURES}
\end{center}

The gravitational vacuum is a gravitational vacuum condensate which has been produced
as the result of Universe creation from "nothing" or as the result of the first
relativistic phase transition. The question is to find the internal structure of
gravitational vacuum starting from quantum regime (Burdyuzha and Vereshkov, 2004).
The quantum regime of gravity has not satisfactory explanation so far although some
analysis of the problem has been undertaken (Carlip, 2001; Banks, 2003).

There is some analogy between the known vacuum structures and a hypothetical structure
of the gravitational vacuum (condensates of the quark-gluon type consist of topological
structures - instantons). Topological microstructures(defects) of gravitational vacuum are:
worm holes, micromembranes, microstrings and a gas of topological monopoles. After Universe
inflation they were smoothed, stretched, and broken up. A part of them could survive and now
they are perceived as the structures of $\Lambda-term$ (quintessence) and the unclustered
dark matter.

The parametrization noninvariance of Wheeler-DeWitt equation can be used to describe
phenomenologically these topological defects. The mathematical illustration of these
processes may be spontaneous breaking of local Lorentz-invariance of the quasi-classical
equations of gravity. The gravitational vacuum condensate has fixed time in our Universe.
Besides, 3-dimensional topological defects (worm-holes) renormalize $\Lambda-term$:
\begin{center}
$$\Lambda = \Lambda_{0} - \frac{\kappa\hbar^{2}c_{3}^{2}}{768\pi^{2}}$$
\end{center}

All parametrizational noninvariant effects are collected in the function $\epsilon_{GVC}(a)$
which we call the density of gravitational vacuum energy and which is proportional
$\sim \hbar^{2}$. It is a clean quantum effect. Probably the parametrizational
noninvariant contributions have arisen because of no conservation of the classical
symmetry on the quantum level. However, general symmetric arguments do not have the clear
physical connection to vacuum energy. In this situation examples of QCD are
useful. In the quantum theory classical conformal and chiral symmetries do not
conserve resulting in appearance of a quark-gluon condensate.

Finally, topological structures exist in gravitational vacuum and they are the consequences
of the parametrizational noninvariance of the quantum geometrodynamics (DeWitt, 1967).
The density of energy of the gravitational vacuum is described by the next formula:

\begin{center}
$$\epsilon_{GVC} = \frac {\kappa\hbar^{2}}{192 \pi^{2} a^{4}}(\mu" - \frac {1}{4}(\mu')^{2})$$
\end{center}
where: $\mu(a) = c_{0} ln a + c_{1} + \frac{1}{2} c_{2} a^{2} + \frac{1}{3} c_{3} a^{3}$ is a parametrizational function;
$c_{i} = const$; and then the sum effect is:
$$\Lambda_{0} + \epsilon_{GVC} (a) = \Lambda_{0} - \frac{\ae \hbar^{2}}
{768 \pi^{2}}\; c^{2}_{3} + \frac{\ae \hbar^{2}}{192 \pi^{2}}
[- \frac{1}{2} \; c_{2} c_{3} \frac{1}{a}$$
$$ - (\frac{1}{4} c^{2}_{2} +
\frac{1}{2} c_{1} c_{3}) \frac{1}{a^{2}} +
(2 c_{3} - \frac{1}{2} c_{1} c_{2} - \frac{1}{2} c_{0} c_{3}) \frac{1}{a^{3}} +
(c_{2} - \frac{1}{4} c^{2}_{1} - \frac{1}{2} c_{0} c_{2}) \frac{1}{a^{4}} $$
$$- \frac{1}{2} c_{0} c_{1} \frac{1}{a^{5}} - (c_{0} + \frac{c^{2}_{0}}{4})
\frac{1}{a^{6}}] $$

The energy density of the system of topological defects contains a constant part
corresponding to worm-holes and members of types $\frac{1}{a^{3}};\frac{1}{a^{2}};\frac{1}{a}$
corresponding to a gas of point defects , micromembranes, microstrings. The term $\frac{1}{a^{3}}$
may be relevant to dark matter also. The last three members can be interpreted as
the energy of gravitational interaction of defects between each other.
\begin{center}
{\bf 8. ROLE OF VACUUM IN THE UNIVERSE}
\end{center}
Probably initial vacuum in the Universe was in a sharply nonequilibrium state and
as the consequence of this was possible inflation. 4-dim was not an initially given
topological configuration. A many dimension Universe might be created. Then
the compactification was a dynamical phenomenon. The creation of the Universe
from "nothing" $(E_{tot} =0)$ is the most probable process from a space-time foam.
Unfortunately the negative moment takes place. We extrapolate the known physical and
logical laws to processes for which they was not formulated. The result of this
extrapolation is the quantum geometrodynamics of the initial Universe evolution.

Note other time that on some reasons (a fluctuation) a closed empty space acquires
some individual geometrical characteristics in which vacuum was in a sharply
nonequilibrium state. The nonequilibrium of vacuum provides the inevitability of
evolution of this empty space. From vacuum particles were created. But the positive
energy of particles and vacuum was exactly compensated the negative energy of gravitation
field (that is $E_{tot}=0$). Thus, without breaking of the law of energy conservation
a primary small Universe has transformed in a large Universe filled matter.

The boundary conditions for a wave function of the Universe choose a closed state without
particles as an initial state (this is the hard physical condition).

The time is the fourth coordinate which is necessary to describe a curved 4-dim Riemann
manifold. Appearance of time in equations of quantum geometrodynamics is
caused by nonsymmetry of vacuum relatively of space-time transformations of Einstein theory.
This effect has been fixed by the gravitational vacuum condensate.

Finally, probably the modern vacuum in the Universe is a small part of initial vacuum
which has left after production of particles in a complex process of vacuum selforganization
dynamical mechanism of which we do not know well till now.
\vspace{2cm}

{\bf REFERENCES}\\
Armendariz-Picon C., Mukhanov V. and Steinhardt P. (2003) Phys.Rev.Lett. 85, 4438.\\
Banks T. (2003) hep-th/0305206.\\
Burdyuzha V. (1998) In Proceedings of PASCOS-98. Ed. World Scientific, p. 101.\\
Burdyuzha V. and Vereshkov G. (2002) Preprint of Lebedev Physical Inst. N 6.\\
Burdyuzha V. and Vereshkov G. (2004) Phys.Rev.D (submitted).\\
Carlip S. Rep. (2001) Prog.Phys. 64, 885.\\
DeWitt B. (1967) Phys.Rev. 160, 1113.\\
Dolgov A. (2004) hep-ph/0405089.\\
Hebecker A. and Wetterich C. (2000) Phys.Rev.Lett. 85, 3339.\\
Kardashev N. (1998) Astron. Zh. 74, 803.\\
Lahav O. and Liddle A.R. (2004) astro-ph/0406681.\\
Latipov N., Beilin V. and Vereshkov G. (2001) "Vacuum, Particles and Universe" Ed.of Moscow University.\\
Riess A.G. et al. (2004) astro-ph/0402512.\\
Rubakov V. (2004) Phys.Rev.D. 61, 061501.\\
Starobinskiy A. (1998) Pis'ma ZhETF 68, 721.\\
Starobinskiy A. (2004) The talk on Moscow astrophysical seminar at 22 Feb.\\
Turok N. and Hawking S. (1998) hep-th/9803156.\\
Vikman A. (2004) astro-ph/0407107.\\
Weinberg S. (1989) Rev.Mod.Phys. 61, 1.\\
Zel'dovich Ya.B. (1967) Pis'ma ZhETP 6, 883.\\

\end{document}